# Void Engineering in Epitaxially Regrown GaAs-Based Photonic Crystal Surface Emitting Lasers by Grating Profile Design

A. F. McKenzie[1,2,a)], B. C. King[1], K. J. Rae[1], S. Thoms[1], N. D. Gerrard[1], J. R. Orchard[2], K. Nishi[3], K. Takemasa[3], M. Sugawara[3], R. J. E. Taylor[1], D. T. D. Childs[1], D. A. MacLaren[4], and R. A. Hogg[1]

[1] School of Engineering, University of Glasgow, Glasgow, G12 8QQ, UK
[2] Compound Semiconductor Technologies Global Ltd., Blantyre, G72 0BN, UK
[3] QD Laser Inc., Minamiwataridacho, Kawaski-ku, Kawasaki, Kanagawa, 210-0855, Japan
[4] SUPA, School of Physics and Astronomy, University of Glasgow, Glasgow, G12 8QQ, UK

[a)]**Authors to whom correspondence should be addressed:** a.mckenzie.1@research.gla.ac.uk

We report the engineering of air-voids embedded in GaAs-based photonic crystal surface emitting lasers realised by metalorganic vapour-phase epitaxy regrowth. Two distinct void geometries are obtained by modifying the photonic crystal grating profile within the reactor prior to regrowth. The mechanism of void formation is inferred from scanning transmission electron microscopy analysis, with the evolution of the growth front illustrated though the use of an AlAs/GaAs superlattice structure. Competition between rapid lateral growth of the (100) surface and slow diffusion across higher index planes is exploited in order to increase void volume, leading to an order of magnitude reduction in threshold current and an increase in output power through an increase in the associated grating coupling strength.

Photonic crystal surface emitting lasers (PCSELs) have emerged as a new class of semiconductor laser that offer single-mode emission with extremely narrow beam divergence and tunable beam shape[1]. These properties are realised by the inclusion of a photonic crystal (PC) grating layer adjacent to the active region of the device, in which a two dimensional variation in refractive index causes the scattering and interference of light and the emergence of a photonic band structure[2]. The scattering condition at the $\Gamma$-point of the band structure results in the formation of a two-dimensional resonant cavity that supports laser oscillation within the plane of the PC, with light scattered out of plane via second-order scattering[3]. Through this band edge resonance effect, light is emitted coherently from the entire surface of the device, promising the possibility of true area-scalable output power without the degradation of optical mode quality.

Whilst originally realised through wafer-bonding[4], metalorganic vapour-phase epitaxy (MOVPE)-based regrowth has emerged as a promising method for PCSEL fabrication owing to the elimination of the interfacial defects associated with bonding. MOVPE regrowth also offers the flexibility to realise both all-semiconductor[5,6] and air-void-based[7,8] PC structures, with watt-class operation of devices having been demonstrated[9]. In this regard, control over the regrowth process is crucial in optimising device performance because properties such as coupling strengths[10] and output power[11,12,13] are dependent on the microscopic geometry of the PC grating unit cell. Although previous studies have investigated the engineering of air-void shape during regrowth[7,14], the scanning electron microscopy (SEM)-based analysis employed revealed little about the evolution of epitaxy, and hence the mechanisms that drive void formation or complete semiconductor infill.

In this work, we present two GaAs-based PCSELs containing voids encapsulated during MOVPE regrowth of a patterned GaAs layer with AlAs/GaAs. The use of an AlAs/GaAs superlattice (SL) provides snapshots of the evolving growth front during regrowth and allows the mechanism of void formation to be inferred by scanning transmission electron microscopy (STEM)-based analysis. The relationship between the crystallographic geometry of the patterned grating and the final void geometry is explored using two distinct grating profiles obtained by exploiting mass-transport and surface restructuring during pre-growth thermal processing within the MOVPE reactor. In particular, the proportion of (100) and the orientation of adjacent facets are shown to be important in influencing the extent of the lateral growth which causes the voids to form. The presence of fast-growing {311}-like planes results in small "pill"-shaped voids. Elimination of these planes from the grating profile produces much larger ellipsoidal voids which lead to significantly improved device performance.

The base-epitaxial wafers for both devices presented in this study were grown by MBE on (100) GaAs with a 2°-offcut towards (1-10). The structure consisted of a thick $n$-AlGaAs cladding layer, an active region composed of three InGaAs/GaAs quantum wells, and a top $p$-GaAs layer. A square-lattice of circular holes with a period of 320 nm was defined in PMMA by electron-beam lithography and patterned into the p-GaAs layer by reactive ion etching (RIE). An SEM image of the as-etched PC is shown in Fig 1(a). In order to maximise the potential coupling strength of the grating[10], each of the holes was etched with a nominal diameter of 256 nm, corresponding to an r/a value (the ratio of the radius to the period) of 0.4. The resulting depth was approximately 150 nm. The total PC area per device was 150 x 150 µm.

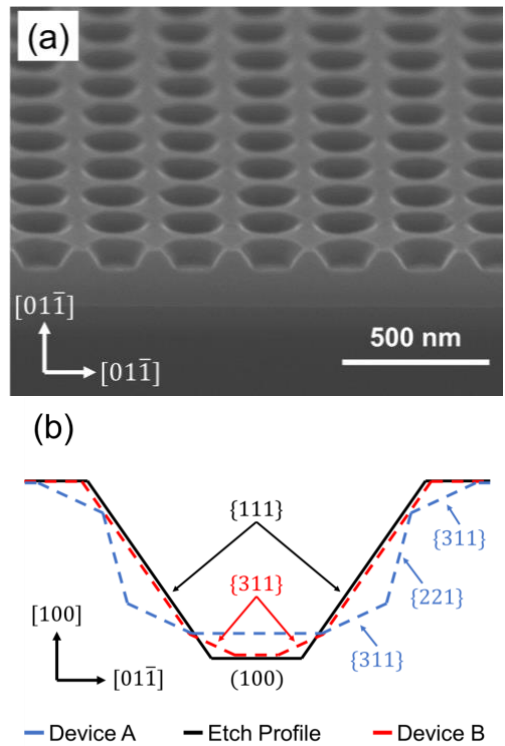

**FIG. 1.** (a) Representative SEM image of a cleaved, as-etched PC layer; viewed along the [011] direction. The circular grating holes have a truncated-cone shape bounded by {111}-like sidewalls. (b) Schematic illustrating the change in grating geometry in Device A and B versus the etched profile.

Immediately prior to MOVPE regrowth, wafers were cleaned using a UVCOS UV/ozone cleaner and etched in 10:1 buffered HF for one minute. Following this, they were loaded into a Thomas Swan 3x2" CCS MOVPE reactor operating at a pressure of 100 mbar. The PC was then regrown with 120 nm of an AlAs/GaAs SL structure at a temperature of 650 °C and V/III ratio of 700, with a nominal growth rate of 6 nm/min on the (100) surface. The SL consists of alternating layers with nominal thicknesses of 9 nm and 1 nm, for AlAs and GaAs, respectively. The PC infill layer was immediately followed by a 1.5 µm thick $p$-doped $Al_{0.41}Ga_{0.59}As$ cladding layer and finished with a 100 nm thick $p^+$-GaAs contact layer. Trimethylaluminium (TMAl) and trimethylgallium (TMGa) were utilized as group-III precursors, with arsine ($AsH_3$) and dimethylzinc (DMZn) as the group-V precursor and dopant source, respectively.

Following MOVPE regrowth, a 100 x 100 µm square mesa, etched to a depth of 300 nm, was defined in the $p^+$-GaAs contact layer using a combination of photolithography and a solution of hydrogen peroxide and orthophosphoric acid. A 200 nm thick $SiO_2$ passivation layer was then deposited across the wafer, into which a 100 x 100 µm contact window was opened on the mesa top using a $CHF_3$/Ar reactive ion etch. A *p*-type Ti/Pt/Au contact was then deposited, and a 60 µm-diameter circular emission aperture defined using a standard lift-off process. Fabrication was completed by the deposition of a backside Ni/Au/Ge/Ni/Au n-type contact and thick Ti/Au bond pads on the top surface.

Whilst the as-etched grating profile was nominally identical in both devices (Fig. 1(a)), the geometry at the start of epitaxy was modified in order to influence the mechanism of void formation. This was realised within the MOVPE reactor by exploiting mass-transport phenomena during the temperature ramp cycle prior to regrowth. Here, the micron-scale gradients in chemical potential that are introduced to the wafer by patterning and the exposure of different crystal planes result in thermally-activated diffusion of Ga as the system attempts to minimise its surface energy[15]. By varying the temperature ramp rate, we can control the extent of Ga mass-transport and modify the grating profile *in-situ*. In this case, the temperature of the wafers was increased linearly from 330 °C to 650 °C under a constant $AsH_3$ flow, ramping over six minutes for Device A and three minutes for Device B.

The difference in cross-sectional profiles is illustrated in Fig. 1(b), where the profiles derive from STEM analysis (see below). Whereas the etched profile (in black) consisted of top and bottom (100) planes connected by {111}-like sidewalls, mass-transport during annealing substantially smooths those seen in the regrown devices. The extent of diffusion increases with anneal time and is greatest for Device A. It is characterised by a reduction in hole depth and overall increase in hole width consistent with mass-conservation in this transport regime. This is accompanied by significant changes in local crystallography within the hole, with the formation of planes adjacent to the top and bottom surfaces and the introduction of short connecting sidewalls. On the basis of the STEM analysis, the angles at which they intersect the (100) plane are consistent with {311} and {221}-like planes, respectively. The reduced ramp time used for Device B, by compassion, resulted in much less diffusion and only minimal smoothing, characterised by a slight reduction in hole depth and the formation of only short {311}-like planes in the bottom corners of the hole. The grating profiles seen in the regrown devices form part of a surface-energy-minimisation sequence that is in good agreement with that predicted by simulations of thermally promoted deformation of comparable pits etched in (100) Si surfaces[16]. Comparisons with such simulations suggest that subsequent deposition of the SL passivates the growth surface to prevent further deformation.

Structural analysis of the regrown PC region was performed using a JEOL ARM200cF STEM operating at 200 kV. Cross-sectional specimens were prepared in a Thermo Fisher Xe-Plasma focussed ion beam (FIB) system by a standard lamella lift-out method[17]. Fig. 2(a) shows a cross-sectional high-angle annular dark-field (HAADF)-STEM image of a single grating period of the PC layer in Device A, as viewed along the [011] direction. As signal intensity in HAADF imaging is proportional to the atomic number squared for each of the elemental species present in the specimen, it follows that the central dark feature is a void that was encapsulated within the AlAs/GaAs infill layer during regrowth. In this case the void is bounded by near-vertical sidewalls giving it a cylindrical "pill"-like shape, with a diameter and height of 50 nm and 120 nm, respectively. Such voids are present above each of the grating features and, due to the considerably larger refractive index difference between semiconductor and "air" void compared to GaAs and AlAs, will relate to the majority of the coupling coefficient for the laser. As such, the presence of the void forms the basis of the PC laser with a nominal r/a value of only 0.08, compared to a value of 0.4 if only the grating geometry is considered.

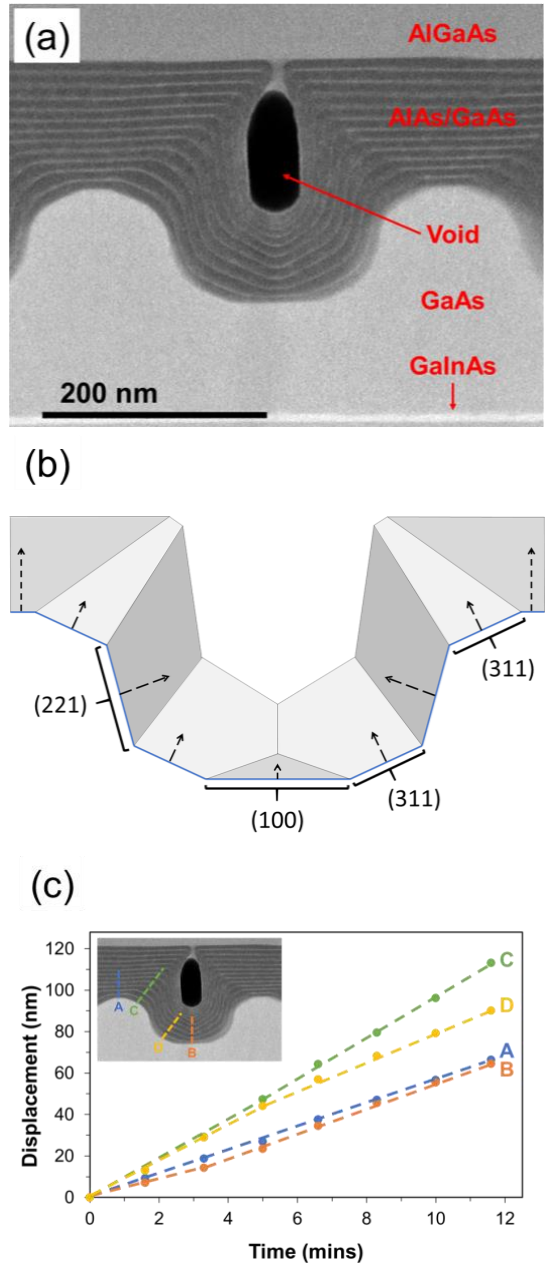

**FIG. 2.** (a) HAADF-STEM image of the regrown PC layer in Device A showing the formation of a void (black). (b) Schematic illustrating the growth front after Deposition of seven SL layers. Cross-sections viewed along the [011] direction. (c) Graph showing the relative growth rates normal to the {100} and {311} planes as measured by the displacement of successive GaAs SL along the lines (A-D) indicated.

The lighter GaAs layers of the SL in the STEM image illustrate the growth front evolution during PC infill, with successive layers providing a time-resolved snapshot as the process proceeds. It is clear from Fig. 2(a) that the growth front evolves through substantial lateral growth of the top (100) plane, with an accompanying reorientation and lengthening of the sidewalls toward the vertical {011} facets which bound the void. This mechanism contrasts with that seen during the growth of V-groove quantum wires[18] and pyramidal quantum dots[19] in which the original etch profile is retained by subsequent layers in what is termed "self-limited" growth. In the case of AlGaAs/GaAs quantum wires, the structures derive from an enhanced growth rate from the trench sidewalls relative to the (100) plane, and allow for the complete planarization of the patterned area without the formation of voids. In contrast, we attribute the void formation seen in Fig. 2(a) to restricted diffusion of absorbed species on the sidewalls, in comparison to the (100) surface, during growth. Diffusion of material off the (100) surface leads to an accumulation of species, and hence apparently thicker SL periods on the slow diffusion bounding facets. It is known that Al has inherently lower mobility compared to GaAs or AlGaAs[20], and lower rates on more open, higher Miller index surfaces is expected. The limited surface mobility is further suppressed by the use of a relatively low growth temperature and large V/III ratio, which have been shown to impede the diffusion of group-III species across similarly patterned substrates[21].

The evolving growth front seen in Device A is illustrated in the cross-sectional schematic shown in Fig. 2(b). This diagram was constructed from the STEM image (Fig. 2(a)) using the seventh GaAs SL layer as a guide to the shape of the growth front after approximately 12 mins of deposition. The front has been simplified to remove the curved interfaces between planes, and a single domain emanating from each of the crystal planes that form the underlying grating (blue line) is highlighted; dashed arrows indicate the surface normal for each plane. Here we see that the lateral growth of the top surface and sidewall re-orientation are accompanied by a shrinking, and eventual elimination, of the top {311} planes. Measurements of the relative growth rates normal to each plane, shown in Fig. 2(c), reveal substantially thicker layers on {311} compared to the adjacent (100). The difference in growth rate between the sets of {311} planes seen in Fig. 2(c) is accentuated by reduced atomic flux impinging on planes at the bottom of the grating hole, resulting in the reorientation and lengthening of the sidewalls. The difference is exacerbated by the low mobility of Al in this system, which prevents the surface diffusion of species to bottom {311} surfaces, and increases over time as the aspect ratio of the aperture through which incident atomic flux passes increases. The void is ultimately formed following the merging of overhanging corners and the elimination of the top {311} planes from the growth front. At this point, lateral growth of the (100) plane dominates over deposition within the large-aspect-ratio hole, allowing the void to be encapsulated without further reduction in diameter.

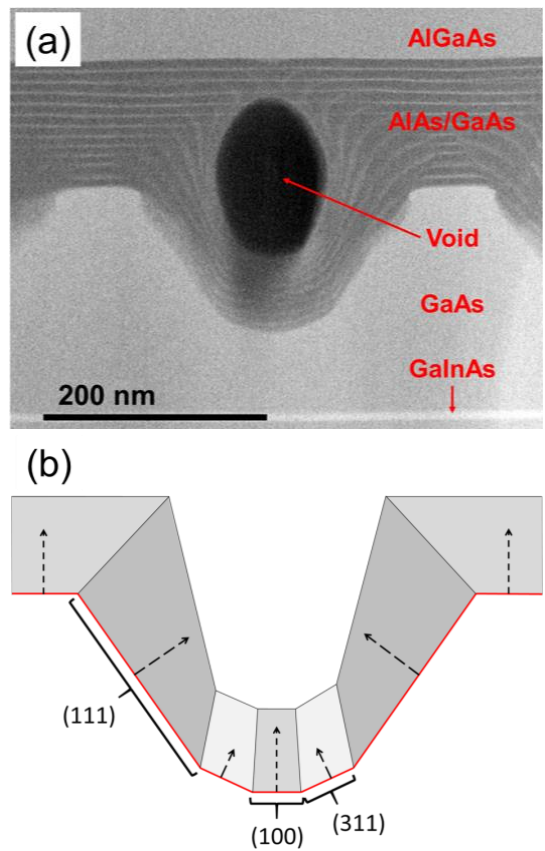

**FIG. 3.** (a) HAADF-STEM image of the regrown PC layer in Device B. (b) Schematic illustrating the growth front after deposition of seven SL periods. Cross-sections viewed along the [011] direction.

A cross-sectional HAADF-STEM image of the regrown PC layer of Device B is shown in Fig. 3(a). Again, regrowth has resulted in void formation however, in this case, the voids are significantly larger than those seen in Device A. Here the void is approximately 155 nm high and is ellipsoidal in shape, with a diameter of 110 nm at its widest point corresponding to an r/a value of 0.17. As discussed above, the 50 % reduction in temperature ramp time compared with Device A was successful in reducing the extent of Ga mass-transport, resulting in the retention of the {111}-like grating sidewalls defined at the etch stage. The evolution of the growth front for this system is illustrated in Fig. 3(b). Exact quantification of the sidewall SL layer thicknesses is limited because the layers appear less distinct, which we attribute to the relative thickness of the STEM cross-section; projection of the three-dimensional structure onto a two-dimensional image. However, there is a clear non-uniformity in the overall thickness, with enhanced growth at the interface of the top (100) plane and the sidewall, and rapid extension of the top (100) surface. This effect is consistent with that seen in the regrowth of V-grooves at lower temperatures and large V/III ratios[21], in which the limited mobility of the group III-species prevents them from diffusing along {111} planes. It is this "bunching" of Al drives the lateral growth of the (100) planes that results in void encapsulation, confirming the importance of suppressed Al mobility during the regrowth process.

The quasi-continuous wave (CW) LI and spectral characteristics of the devices were measured at 20 °C using a pulse width of 2 µs and 1 % duty cycle. In the case of Device A, lasing occurred at a threshold current of 440 mA (J = 4.4 kA/cm$^2$) and a wavelength of 1074 nm. In addition, the device suffered from a low slope efficiency with an output power of only 64 pW recorded at 600 mA. The poor performance of the device can be attributed to the size of the PC voids. Previous simulations of square-lattice PCs consisting of circular holes have suggested that local maxima in the in-plane coupling coefficient are obtained when the r/a is either 0.2 and 0.4[10]. The sub-optimal r/a of 0.07 for the voids in Device A are therefore associated with large values of in-plane loss and, in turn, large threshold current and low slope efficiency for the vertically emitted light. Similarly, it has been shown that, for a given r/a, maximum output power is achieved when the void height is equal to half of the grating period[11]. In this case the height is only 75 % of the optimum value of 160 nm, and so the output power is further diminished due to destructive interference of light scattered out of plane.

Device B displayed significantly improved laser performance over A, with an order of magnitude reduction in threshold current to 65 mA (J = 0.65 kA/cm$^2$), and a measured output power of 3.61 mW at 300 mA, as shown in Fig. 4. The associated slope efficiency is 0.012 W/A, however it should be noted that light is collected from an aperture of only 28 % of the total pumped area due to obscuration by the top gold contact. The optical spectrum of the device (Fig. 4, inset) is characterised by a main lasing peak centred at 1074 nm, with a less intense peak at a shorter wavelength owing to the narrow splitting of bands at the $\Gamma_2$-point of the photonic band structure. The improvements in threshold and output power can be attributed to the increase in grating coupling strength and decrease in vertical destructive interference associated with improved void radius (0.17 r/a) and height (155 nm), respectively, in line with the predictions of the simulation-based studies mentioned above[10,11].

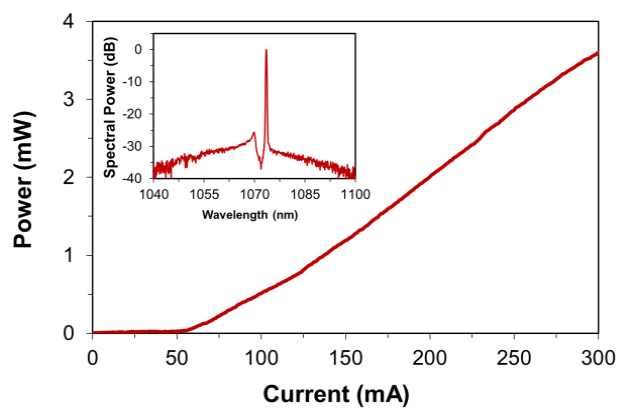

**FIG. 4.** Quasi-CW LI and lasing spectrum (inset) of Device B measured at 20 °C with a pulse width of 2 µs and 1 % duty cycle.

We have presented a study of growth-front development in MOVPE-regrown GaAs-based PCSELs containing air-void PC, made possible by the use of an AlAs/GaAs superlattice structure. Two unique void geometries have been realised by controlling the extent of mass-transport-driven deformation of the PC grating during the pre-growth temperature ramp process within the reactor. STEM analysis of the regrown devices reveals the evolution of the growth front in each case, confirming that void formation is the result of rapid lateral growth of the inter-hole (100) surface and slow diffusion across higher order surfaces binding the hole. The device containing voids with a larger volume exhibited greatly improved threshold current and output power, confirming the results of previous simulation-based studies. Further optimisation of void geometry and device parameters can be achieved by considering different growth temperatures and V/III ratio.

This work was supported by the Engineering and Physical Sciences Research Council (Grant No. EP/S023321/1). AFM is grateful for support from CST Global Ltd. through an Industrial Fellowship from the Royal Commission for the Exhibition of 1851. The data that support the findings of this study are available from the corresponding author upon reasonable request.